# The neutrino signal from protoneutron star accretion and black hole formation

T. Fischer[1], S. C. Whitehouse[1], A. Mezzacappa[2], F.-K. Thielemann[1] and M. Liebendörfer[1]

[1] Department of Physics, University of Basel, Klingelbergstrasse 82, 4056 Basel, Switzerland
   *email* tobias.fischer@unibas.ch
[2] Physics Division, Oak Ridge National Laboratory, Oak Ridge, Tennessee 37831-1200

**Abstract**

*Context.* We discuss the formation of stellar mass black holes via protoneutron star (PNS) collapse. In the absence of an earlier explosion, the PNS collapses to a black hole due to the continued mass accretion onto the PNS. We present an analysis of the emitted neutrino spectra of all three flavors during the PNS contraction.
*Aims.* Special attention is given to the physical conditions which depend on the input physics, e.g. the equation of state (EoS) and the progenitor model.
*Methods.* The PNSs are modeled as the central object in core collapse simulations using general relativistic three-flavor Boltzmann neutrino transport in spherical symmetry. The simulations are launched from several massive progenitors of 40 $M_\odot$ and 50 $M_\odot$.
*Results.* We analyze the electron-neutrino luminosity dependencies and construct a simple approximation for the electron-neutrino luminosity, which depends only on the physical conditions at the electron-neutrinosphere. In addition, we analyze different $(\mu,\tau)$-neutrino pair-reactions separately and compare the differences during the post-bounce phase of failed core collapse supernova explosions of massive progenitors. We also investigate the connection between the increasing $(\mu,\tau)$-neutrino luminosity and the PNS contraction during the accretion phase before black hole formation.
*Conclusions.* Comparing the different post bounce phase of the progenitor models under investigation, we find large differences in the emitted neutrino spectra. These differences and the analysis of the electron-neutrino luminosity indicate a strong progenitor model dependency of the emitted neutrino signal.

## 1. Introduction

Core collapse supernovae of progenitor stars in the mass range of 8 ~ 75 $M_\odot$ release energy of several $10^{53}$ erg/s in neutrinos on a timescale of seconds. These neutrinos carry detailed information from the interior of the stellar core and are therefore of special interest for neutrino measurement facilities, such as Super Kamiokande and SNO that are able to detect a Galactic core collapse supernova at high resolution. In addition, gravitational waves, nucleosynthesis yields and neutron star (NS) properties are able to provide information about the physical conditions and the dynamical evolution inside the stellar core. Up to now gravitational waves have proven difficult to detect and nucleosynthesis calculations are model dependent. Apart from NS mass measurements, which are able to provide constraints on the equation of state (EoS) for hot and dense nuclear matter, neutrinos are the most promising source of information leaving collapsing stellar cores. Core collapse supernova modelers are now yearning for a Galactic explosion, to be able to compare the theoretically predicted neutrino signal from simulations with the actual measured one. So far SN1987A produced the only measured neutrino signal from a core collapse supernova event, which while providing very few data points, nevertheless enables scientists to probe the theoretically predicted scenario to a limited extent (see Hirata et al. (1988)).

At the end of nuclear burning, progenitors stars more massive than 8 $M_\odot$ collapse due to the pressure loss via photodisintegration of heavy nuclei and electron captures. The central density increases beyond nuclear saturation density of $\rho \simeq 2 \times 10^{14}$ g/cm$^3$ (depending on the equation of state EoS). The incompressibility increases and the collapse halts. The core bounces back and a sound wave forms, which turns into a shock front meeting the supersonically infalling material at the sonic point. The electron-neutrinos, which are emitted via additional electron captures after bounce, leave the star in the so called neutronization burst within a few milliseconds after bounce as soon as the shock propagates into the density regime below neutrino trapping of $\rho \simeq 10^{12}$ g/cm$^3$. This energy loss, in combination with the dissociation of infalling heavy nuclei, quickly turn the shock into a standing accretion shock (SAS), which expands to a few 100 km at about 50 ms after bounce. The dissociated nucleons behind the shock accrete slowly onto the central object, a protoneutron star (PNS) of initial radius of $\simeq 50$ km, which contracts on timescales depending on the accretion rate and the assumed matter conditions of the infalling matter.

On the other hand, neutrino reactions behind and ahead of the SAS have long been investigated as a possible source of energy leading to so called neutrino driven explosions [Bethe and Wilson (1985), Janka (2001), Janka et al. (2005), Mezzacappa et al. (2006)]. However, the inefficiency of the neutrino heating and the absence of any explosions using spherically symmetric core collapse simulations of progenitors more massive than the 8.8 $M_\odot$ O-Ne-Mg-core [see Kitaura et al. (2006)] from Nomoto (1983, 1984, 1987), implies the missing of (some) important ingredient(s), most likely multi-dimensional phenomena such as convection, rotation and the development of fluid instabilities.

Multi-dimensional core collapse models have become available only recently [Herant et al. (1994), Burrows et al. (1995), Buras et al. (2003), Bruenn et al. (2006), Buras et al. (2006), Scheck et al. (2008), ] and demonstrate the complexity of the underlying physical phenomena. Due to computational limits, such models have to make use of a neutrino transport approximation scheme. An exception is the axially symmetric core collapse model from Marek and Janka (2007) (and references therein), where Boltzmann neutrino transport is calculated in angular rays. However, the results achieved via these mostly axially symmetric core collapse models will have to be confirmed by three dimensional models. On the other hand, although spherically symmetric core collapse models fail to explain the explosion mechanism of massive progenitor stars, such models are well suited for investigating accretion phenomena and the emitted neutrino spectra accurately up to several seconds after bounce. Such long simulation times are at present beyond the numerical capability of multi-dimensional models.

The present paper reports on the emitted neutrino signal from failed core collapse supernova explosions and the formation of black holes via PNS collapse [see for example Baumgarte et al. (1996), Beacom et al. (2001), Liebendörfer et al. (2004), Sumiyoshi et al. (2007)]. We perform general relativistic simulations in spherical symmetry using spectral three-flavor Boltzmann neutrino transport. By our choice of a spherically symmetric approach, we assume that accurate neutrino transport and general relativistic effects are more important for the analysis of the emitted neutrino signal, than multi-dimensional phenomena which are investigated in Marek et al. (2008). The simulations are launched from several massive progenitors stars of 40 and 50 $M_\odot$. For such progenitors, the presence of strong gravitational fields implements that general relativistic effects are important and must be taken into account.

Such progenitor stars will not develop explosions but collapse to a black hole, while low and intermediate mass progenitors are assumed to explode and leave a NS behind. After the explosion has been launched, the NSs contract on timescales of millions of years and cools in several phases. [Page (1995), Pons et al. (2002), Henderson and Page (2007)]. Progenitor stars in the mass range between intermediate mass and $\simeq 40$ $M_\odot$ are assumed to explode as well but the material that falls back onto the PNS may force the PNS to collapse to a solar mass black hole. Neutron pressure and nuclear forces will eventually fail to keep the PNS stable against gravity. The actual progenitor mass limits for these different scenarios are a subject of debate and depend on the explosion mechanism and the input physics. The fate of progenitors more massive than $\simeq 75$ $M_\odot$ differs from the usual Fe-core collapse scenario by the appearance of the pair-instability, as studied by Fryer et al. (2001), Linke et al. (2001), Heger and Woosley (2002), Nomoto et al. (2005), Ohkubo et al. (2006) and Nakazato et al. (2007).

In this paper, we will investigate the differences in the emitted neutrino signals for several massive progenitor models from different stellar evolution groups [Woosley and Weaver (1995), Heger and Woosley (2002), Umeda and Nomoto (2008), Tominaga et al. (2007)] during the accretion phase before black hole formation.

The emitted neutrino signal depends on the matter conditions during the dynamical evolution of the PNS contraction, which are given by the EoS for hot and dense nuclear matter. Sumiyoshi et al. (2007) compared two different EoSs with respect to stiffness and illustrated the different emitted neutrino signals, especially the different timescales for the PNSs to become gravitationally unstable during the accretion phase of failed core collapse supernova explosions. We will point out the importance of the progenitor model for the emitted neutrino signal. We will demonstrate that it is hardly possible to draw any conclusions from the neutrino signal about the EoS or the progenitor model separately, as both quantities have similar effects on the emitted neutrino spectra.

Moreover, we analyze the feasibility of approximating the electron-neutrino luminosity at large distances (typically of the order of a few 100 km and more) depending explicitly on the progenitor model (the mass accretion rate) and the temperature at the neutrinosphere. Liebendörfer (2005) presented a density-parametrized deleptonization scheme which can be applied in multi-dimensional simulations during the collapse phase. Here, we introduce a simple model to illustrate the dependency of the electron-(anti)neutrino luminosity from the matter conditions at the PNS surface.

$(\mu/\tau)$-(anti)neutrinos are assumed to interact via neutral current reactions only as the thermodynamic conditions do not favor the presence of a large fraction of $\mu/\tau$. Hence the muonic charged current reactions are suppressed. On the other hand, the $(\mu/\tau)$-(anti)neutrino emission is rather important for the cooling at the $(\mu/\tau)$-neutrinospheres and needs to be handled carefully. In an earlier study, Liebendörfer et al. (2004) emphasized the $(\mu/\tau)$-(anti)neutrino luminosity increase during the accretion phase of a 40 $M_\odot$ progenitor model. Fischer et al. (2007) extended this study and presented preliminary results investigating the connection between the $(\mu/\tau)$-(anti)neutrino luminosity increase and the contraction of the PNS during the accretion phase. Here, we will compare selected pair creation reactions separately and analyze the consequences of these different reactions to the post bounce evolution of massive progenitors before black hole formation.

The paper is organized as follows. In §2, we summarize our spherically symmetric core collapse model and compare the recently implemented EoS from Shen et al. (1998) with the EoS from Lattimer and Swesty (1991) using the example of a core collapse simulation of a 40 $M_\odot$ progenitor. In addition, we will illustrate the general scenario of the PNS collapse to a black hole. In §3, we introduce the electron neutrino luminosity approximation and apply it to core collapse simulations of several massive progenitors of 40 and 50 $M_\odot$ and an intermediate mass progenitor of 15 $M_\odot$. For the massive progenitor models, we compare the different emitted neutrino signals in §4. We also analyze the different massive progenitor models at the final stage of stellar evolution and investigate the connection between the mass accretion rate at the PNS surface and the progenitor structure. In §5, we discuss different $(\mu/\tau)$-(anti)neutrino reactions and the connection between the PNS contraction and the $(\mu/\tau)$-(anti)neutrino luminosity increase during the post bounce accretion phase of massive progenitors. We also illustrate the corrections of the standard opacities with respect to weak magnetism, nucleon-recoil and contributions from strange quarks as suggested by Horowitz (2002).

## 2. GR radiation hydrodynamics in spherical symmetry

Our model is based on an implicit three-flavor neutrino and anti-neutrino Boltzmann transport solver developed by Mezzacappa & Bruenn (1993a-c) (for a detailed overview on the neutrino physics, see Schinder and Shapiro (1982), Bruenn (1985), Mezzacappa and Messer (1999)). In order to treat the post-bounce phase, Liebendörfer et al. (2001a) coupled this Lagrangian model to an implicit general relativistic hydrodynamics code that features an adaptive grid. In addition, the Boltzmann solver has been extended to solve the general relativistic transport equation described by Lindquist (1966). Liebendörfer et al. (2001b) and Liebendörfer et al. (2004) implemented a finite differencing of the coupled transport and hydrodynamics equations that accurately conserves lepton number and energy in the post-bounce phase.

### 2.1. Aspects of PNS evolution and black hole formation

Fig. 1 illustrates the physical conditions during the collapse of a PNS to a black hole. The PNSs are modeled as the central object in failed core collapse supernova explosions of massive progenitors of (40 and 50 $M_\odot$). As the central density in Fig. 1 (b) exceeds a certain critical value (depending on the EoS), nuclear forces and neutron pressure fail to keep the PNS stable against gravity and the central part of the PNS starts to contract. This can be identified at the radial velocities in Fig. 1 (a). During the subsequent compression, the central matter density in Fig. 1 (b) continues to rise above $10^{15}$ g/cm$^3$ while the shock position remains almost unaffected at $30 - 35$ km. In addition, the hydrodynamical timescale for the PNS to become gravitationally unstable and to collapse to a black hole is reduced to $10^{-6}$ s.

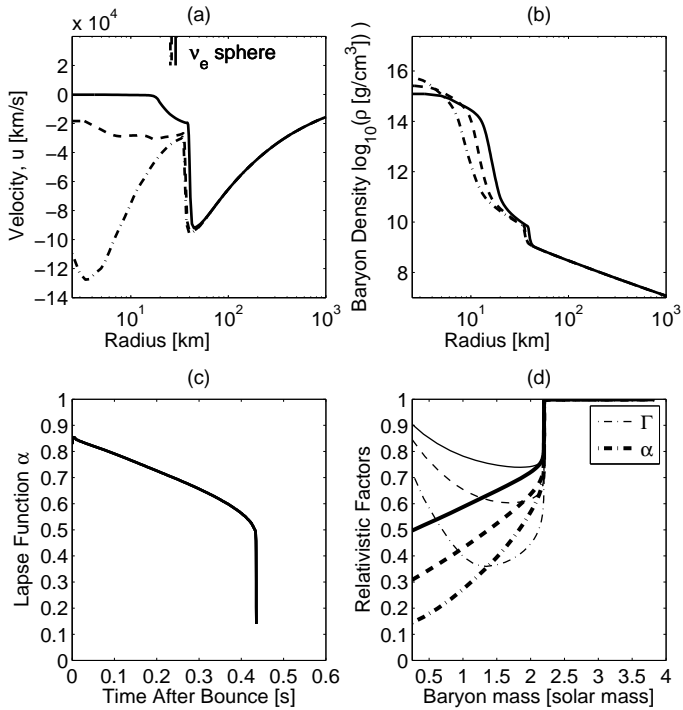

**Figure 1.** Radial velocity and density profiles as a function of the radius. The relativistic factor $\Gamma$ and the lapse function $\alpha$ as a function of the enclosed baryon mass at three different times after bounce during the PNS collapse (solid line 433.6 ms, dashed 435.4 ms, dash-dotted 435.5 ms), at the example of a 40 $M_\odot$ progenitor from Woosley and Weaver (1995). In addition, the lapse function at the center is plotted with respect to time after bounce.

The moment of black hole formation is reached when the central lapse function ($\alpha$ in Fig. 1 (c)) approaches zero with respect to time. Due to our co-moving coordinate choice, no stable solutions of the differential equations for momentum and energy conservation

$$T^{\mu\nu}{}_{;\nu} = 0, \quad (1)$$

with

$$T^{tt} = \rho(1 + e + J), \quad (2)$$
$$T^{aa} = p + \rho K, \quad (3)$$
$$T^{ta} = T^{at} = \rho H, \quad (4)$$
$$T^{\vartheta\vartheta} = T^{\varphi\varphi} = p + \frac{1}{2}\rho(J - K), \quad (5)$$

can be found. $\rho$, $e$ and $p$ are the matter density, internal energy density and pressure respectively. $J, H, K$ are the neutrino momenta which are the phase space integrated neutrino distribution functions given in Liebendörfer et al. (2004). The differential equation for the lapse function $\alpha(t, a)$ follows directly from the Einstein-equation

$$G_{\mu\nu} = \kappa T_{\mu\nu},$$

with the Einstein tensor $G_{ik}$ on the left hand side and the gravitational constant $\kappa$ and stress-energy tensor on the right hand side, for a spherically symmetric and non-stationary spacetime

$$ds^2 = \alpha^2 dt^2 + \left(\frac{r'}{\Gamma}\right)^2 da^2 + r(t,a)^2 \left(d\vartheta^2 + \sin^2\vartheta d\varphi^2\right), \quad (6)$$

(see Misner and Sharp (1964)) with coordinate choice $(t, a)$ eigentime and enclosed baryon mass respectively and the angles $(\vartheta, \varphi)$ describing the 2-sphere. For a detailed derivation of the general relativistic equations used, see Liebendörfer et al. (2001b) and Liebendörfer et al. (2004).

In addition, Fig. 1 (d) shows the relativistic factor $\Gamma = \sqrt{1 - 2m/r + u^2}$ and $\alpha$ as a function of the enclosed baryon mass $a$ for three snapshots directly before and during the PNS collapse, illustrating the region where relativistic effects become important.

### 2.2. The equation of state

At the present time, the EoS for hot and dense asymmetric nuclear matter in core collapse supernovae is uncertain and a subject of research. The information leaving the central core of collapsing stars is limited as matter is optically opaque and the analysis of the escaping neutrinos is difficult, taking into account the emission, absorption, transport and possible oscillations of neutrinos. Hence, the EoSs of present core collapse supernova models have to rely on nuclear physics calculations that are gauged to reproduce data from terrestrial experiments. Depending on the nuclear physics model and the parameters used, the resulting high-density EoSs may differ quite a lot and the differences between core collapse simulations using these EoSs can be large as well.

The EoS has to handle several different regimes, coupled sensitively to each other. At low matter density and temperature ($T < 0.5$ MeV), nuclei and nuclear burning processes dominate the evolution of physical quantities, for instance internal energy, pressure, entropy and the neutron and proton chemical potentials. To handle this regime, we use an ideal gas approximation to calculate the internal energy (coupled to a nuclear burning energy approximation; for simplicity we assume $^{28}$Si and $^{30}$Si here), which is sufficient for the outer layers of the innermost $3 - 4$ $M_\odot$ of 40 and 50 $M_\odot$ progenitors). For low and intermediate mass progenitors, a nuclear reaction network is applied to calculate the energy exchange rate from a $\alpha$-network. However, the low density and low temperature regime is coupled to the EoS for hot and dense nuclear matter above temperatures of $T = 0.5$ MeV, where nuclei are assumed to be in nuclear statistical equilibrium (NSE). In addition, there are free nucleons and light nuclei. The transition from NSE to bulk nuclear matter (free nucleons only) above the neutron drip line is handled via the EoSs intrinsically.

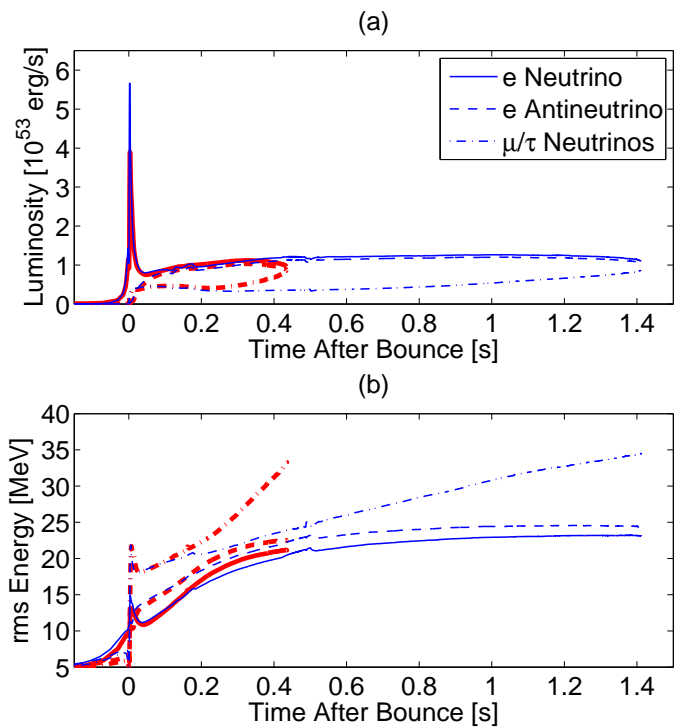

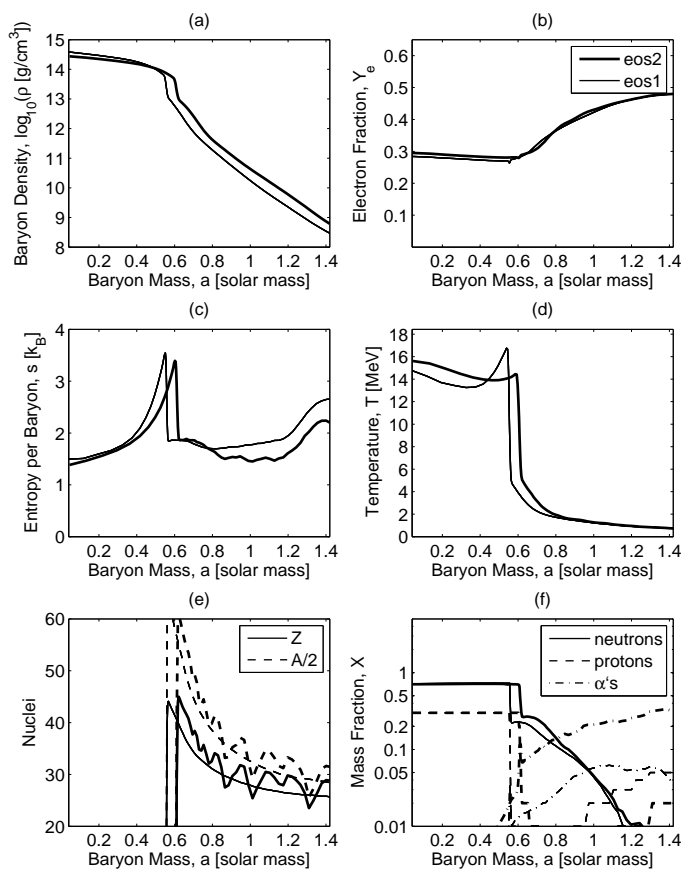

**Figure 2.** Luminosities and mean energies during the post bounce phase of a core collapse simulation of a 40 $M_\odot$ progenitor model from Woosley and Weaver (1995). Comparing *eos1* (thick lines) and *eos2* (thin lines).

**Figure 3.** Bounce conditions for the core collapse simulation of a 40 $M_\odot$ progenitor model from Woosley and Weaver (1995), comparing *eos1* (thin lines) with *eos2* (thick lines).

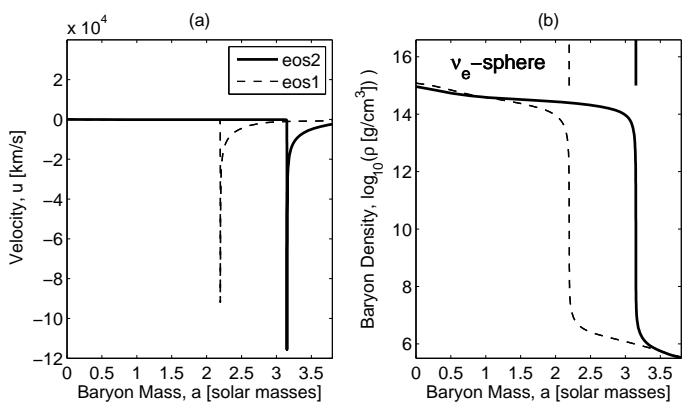

**Figure 4.** The last stable configuration of the PNSs before becoming gravitationally unstable and collapsing to a black hole, comparing *eos1* (thin dashed lines) and *eos2* (thick solid lines).

In the following paragraphs, we compare the soft EoS from Lattimer and Swesty (1991) (*eos1*) with the compressibility of 180 MeV with the stiff EoS from Shen et al. (1998) (*eos2*) with the compressibility of 281 MeV during the accretion phase of a core collapse simulation of the 40 $M_\odot$ progenitor model from Woosley and Weaver (1995) before black hole formation.

*Eos1* is based on the liquid drop model including surface effects, while *eos2* uses a relativistic mean field approach and Thomas-Fermi approximation. In addition, the nuclear part of *eos2*, given as a table, is coupled to an electron-positron EoS, developed by Timmes and Arnett (1999) and Timmes and Swesty (2000). *Eos1*, distributed as a subroutine, already contains the electron-positron contributions. Both EoSs depend on the three independent variables temperature, electron fraction and matter density.

**Table 1.** Thermodynamic conditions of the PNSs in Fig. 4, comparing the central data (a) with the maximum temperature (b).

| EoS | T [MeV] | $\rho$ $10^{14}$ [g/cm$^3$] | $Y_e$ |
|---|---|---|---|
| *eos1* (a) | 29.78 | 12.0 | 0.299 |
| *eos1* (b) | 96.05 | 5.86 | 0.270 |
| *eos2* (a) | 50.27 | 9.16 | 0.297 |
| *eos2* (b) | 92.04 | 4.37 | 0.271 |

Fig. 2 compares the neutrino luminosities in graph (a) and the mean neutrino energies in graph (b) as a function of time after bounce for *eos1* and *eos2*. The larger electron-neutrino luminosity slightly before and at bounce is due to the different thermodynamic conditions achieved at bounce as illustrated in Fig. 3. These different conditions are a direct hydrodynamic consequence of the more compact bouncing core using *eos1* (see the higher central density in graph (a)), which results in a larger central deleptonization in graph (b). The corresponding entropy and temperature profiles are shown in graphs (c) and (d) respectively. At intermediate densities and temperatures, heavy nuclei appear with slightly larger average atomic charge and number using *eos1* (see graph (e)). On the other hand, the fractions of light nuclei in graph (f) differ quite a lot. During the postbounce phase, the simulation using the soft EoS *eos1* is characterized by a short accretion time of $\simeq 500$ ms and thus a rapid PNS contraction before becoming gravitationally unstable and collapsing to a black hole.

Fig. 4 illustrates the last stable configuration before the PNSs (identified via the $\nu_e$-spheres) become gravitationally unstable. Graphs (a) and (b) compare the velocity and the density profiles respectively with respect to the enclosed baryon mass.

The configuration achieved using the stiff *eos2* is supported via larger pressure and nuclear forces, which stabilize the PNS against gravity and allow more mass to be accreted. The maximal masses for both (hot and dense) EoSs were found to be 2.196 M$_\odot$ for *eos1* and 3.15 M$_\odot$ for *eos2* respectively. This results in an extended PNS accretion phase of $\simeq 1.4$ s using *eos2* for the progenitor model under investigation. The corresponding thermodynamic conditions for the PNS configurations illustrated in Fig. 4 are shown in Tab. 1. (a) compares the central data with (b) the maximum temperatures achieved, illustrating the region where the PNSs become gravitationally unstable and start to collapse. However, both of the PNS collapses to a black hole proceed along similar paths.

Our results are in qualitative agreement with an independent study on the subject of an EoS comparison, recently published by Sumiyoshi et al. (2007).

## 3. A simple model for the electron-(anti)neutrino luminosity

Since the core of massive stars is optically opaque, the only sources of information that is able to leave are gravitational waves and neutrinos. An indirect insight into the happenings inside the Fe-core is given by the observed composition of the ejecta in the case of an explosion. However, gravitational waves have proven difficult to detect and nucleosynthesis calculations are model dependent. Neutrinos on the other hand (especially the electron-flavor neutrinos), are of interest for neutrino detector facilities, such as Super-Kamiokande and SNO, being able to resolve the neutrino signal from a Galactic core collapse supernova on a tens of millisecond timescale. The understanding and modeling of the neutrino emission, absorption and transport is essential in core collapse supernova models to be able to compare the predicted neutrino signal with a possible future measurement.

For the prediction of three flavor neutrino spectra, accurate Boltzmann neutrino transport can only be applied in spherically symmetric core collapse simulations. Due to computational limits, multi-dimensional models (especially 3-dimensional models) have to make use of some neutrino transport approximation scheme. An exception is the ray-by-ray discretization in the two-dimensional core collapse model from Marek and Janka (2007), where Boltzmann neutrino transport is calculated in each ray separately. This technique is computationally very expensive. A very simple but powerful approximation has been published by Liebendörfer (2005), applying a density parametrization of the deleptonization during the collapse phase. Unfortunately, it does not reproduce the additional deleptonization after bounce (neutronization burst). With special focus on multi-dimensional simulations of the post bounce phase, we present an analysis of the electron-neutrino luminosity. We will construct an electron-neutrino luminosity approximation, which depends only on the physical conditions at the electron neutrinosphere $R_{\nu_e}$ and can be applied after the neutronization burst after bounce has been launched. We will also compare the approximation with spherically symmetric simulations using three-flavor Boltzmann neutrino transport.

The long term sources of energy for the electron neutrino luminosity are the total change of the potential energy given by the amount of accreted mass per unit time (accretion luminosity)

$$L_{\dot{M}} = \frac{GM}{r}\dot{M}, \qquad (7)$$

at $R_{\nu_e}$. Short term neutrino emission depends on the temperature increase at the neutrinosphere (diffusion luminosity)

$$L_D = 4\pi R_{\nu_e}^2 c\, u_{\nu_e}. \qquad (8)$$

Here, $u_{\nu_e} \propto T^4$ is the thermal black body energy spectrum for ultra-relativistic fermions with matter temperature $T$. Hence, $L_D$ depends only on the thermodynamic conditions at the electron-neutrinosphere, which are given by the PNS contraction behavior and is correlated to the mass accretion rate.

The accurate neutrino number density from Boltzmann transport is (in spherical symmetry) given by

$$\langle n_{\nu_e} \rangle = \frac{4\pi}{(hc)^3} \int_0^\infty E^2 dE \int_{-1}^{+1} \mu d\mu f_{\nu_e}(t, a, \mu, E), \qquad (9)$$

at $R_{\nu_e}$. $f_{\nu_e}(t, a, \mu, E)$ is the electron-neutrino distribution function, which depends (in spherical symmetry) on the phase space coordinates $(t, a, \mu, E)$ where $\mu = \cos\theta$ is the cosine of the neutrino propagation angle $\theta$ and $E$ is the neutrino energy.

We have found that the assumption of a thermal electron-neutrino number spectrum at $R_{\nu_e}$,

$$\langle n_{\nu_e} \rangle \equiv n_\nu \propto T^3,$$

does not generally apply for all progenitor models. The measure of deviation is denoted as

$$\beta = \frac{\langle n_\nu \rangle}{n_\nu}. \qquad (10)$$

Comparing the electron-neutrino luminosity from Boltzmann transport calculations (at large distances; typically of the order of a few 100 km or more) with the luminosities given by Eq. (7) and Eq. (8), we find the following approximation for the electron-neutrino luminosity

$$L_{\nu_e} = \min\left(\frac{1}{4}L_D, \beta L_{\dot{M}}\right). \qquad (11)$$

The pre-factor 1/4 is in agreement with Janka (2001) (see §6.1) and expresses the approximate amount of outward directed transported thermal neutrinos. These neutrinos are assumed to carry information about the local thermodynamic matter conditions, since the neutrino temperature can be approximated by the matter temperature at $R_{\nu_e}$.

Due to the lack of the neutrino momenta in approximate neutrino transport calculations, it is usually not possible to calculate the coefficient $\beta$. However, since $\beta$ depends only on the mass accretion induced deviation of the neutrino spectrum from a black body spectrum, an exponential behavior of the quotient of the accretion luminosity and the diffusion luminosity was empirically found

$$\beta \simeq \frac{1}{2} e^{\frac{L_{\dot{M}}}{L_D}}. \qquad (12)$$

$\beta$ can be understood as a function of the temperature scaled with $\dot{M}$, since the temperature at $R_{\nu_e}$ is adjusted by the PNS contraction given by the mass accretion rate.

In the following section, we will compare the approximation Eq. (11) with three-flavor Boltzmann neutrino transport during the post-bounce evolution of different progenitor models.

### 3.1. Thermal electron-neutrino spectra

The large mass accretion rates of $\simeq 1 - 2$ M$_\odot$/s of the 40 and 50 M$_\odot$ progenitor models from Woosley and Weaver (1995) in Fig. 5 and from Tominaga et al. (2007); Umeda and Nomoto (2005) in Fig. 6 respectively (see graphs (c)) result in fast contracting PNSs (see graphs (d)). The electron-neutrino number densities differ only slightly from a thermal spectrum after the neutrino burst has been launched after about 50 ms post bounce. Hence, the electron-neutrino luminosities at large distances (here 500 km) in the graphs (b) are dominated by the diffusion luminosity over the accretion luminosity due to the limiter in Eq. (11). Graphs (a) compare $\beta$ from Boltzmann transport calculations via Eq. (10) and via Eq. (12). $\beta$ were found to be 0.7, increasing after 100 ms after bounce up to 0.8.

Finally, the fast contracting PNSs become gravitationally unstable rather quickly (due to the soft EoS *eos1* and the large mass accretion rate) and collapse to black holes after 435.5 ms after bounce for the 40 M$_\odot$ progenitor model and after 487.3 ms after bounce for the 50 M$_\odot$ progenitor model.

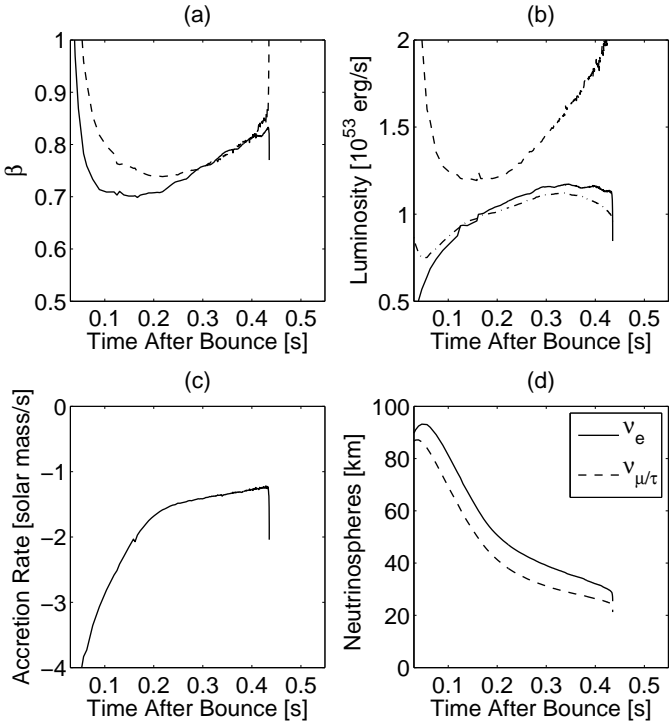

**Figure 5.** The electron-neutrino luminosity approximation and Boltzmann neutrino transport calculations during the post bounce evolution of the 40 M$_\odot$ progenitor from Woosley and Weaver (1995). Graph (a) compares $\beta$ from simulations with Boltzmann neutrino transport Eq. (10) (dashed line) with Eq. (12) (solid line) and graph (b) illustrates the different luminosities separately (solid line: diffusion part of Eq. (11), dashed: accretion part of Eq. (11), dash-dotted: Boltzmann neutrino transport). In addition, graph (c) shows the mass accretion rate at the radius of the electron-neutrinosphere and graph (d) shows the contracting electron- and ($\mu/\tau$)-neutrinospheres.

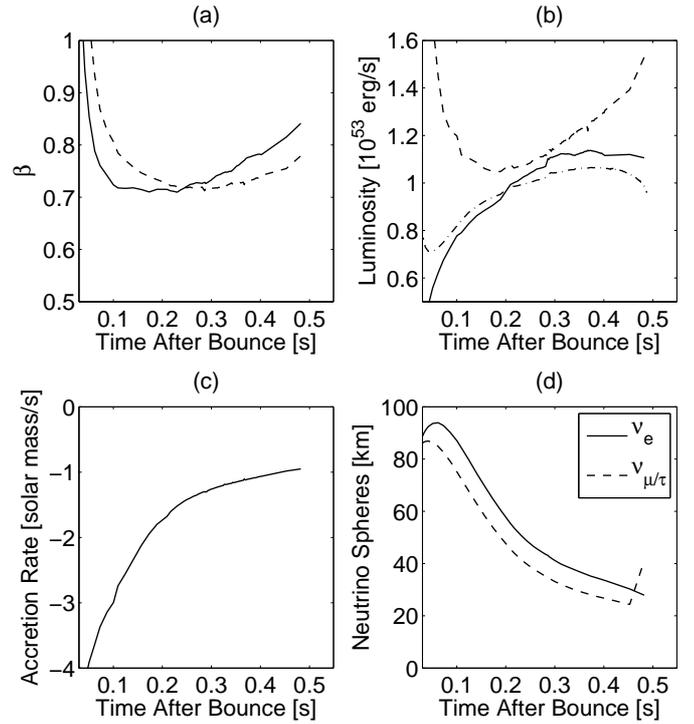

**Figure 6.** The electron-neutrino luminosity approximation and Boltzmann neutrino transport calculations during the post bounce evolution of a 50 M$_\odot$ progenitor from Tominaga et al. (2007). The same presentation of quantities as Fig. 5.

### 3.2. Non-thermal electron-neutrino spectra

We will continue the analysis from above and present data from core collapse simulations of massive progenitors with small mass accretion rates. These models show a different electron-neutrino luminosity dependency with respect to the approximation Eq. (11) during the post bounce phase. Fig. 7 and Fig. 8 illustrate the post bounce evolution of a 40 M$_\odot$ progenitor model from Woosley et al. (2002) and a 50 M$_\odot$ progenitor model from Umeda and Nomoto (2008) respectively (both using *eos1*).

We find the electron neutrino luminosities are initially (until $\simeq 150$ ms post bounce) dominated by the diffusion approximation of Eq. (11), as the matter temperatures are moderately high. This is in agreement with an earlier study by Liebendörfer et al. (2004). However, as the accretion rates in Fig. 7 and Fig. 8 graphs (c) decrease drastically after $\simeq 150$ ms post bounce (even below 0.5 M$_\odot$/s). The temperature at $R_{\nu_e}$ increases less rapidly and the neutrino number density at $R_{\nu_e}$ differs from a thermal spectrum. $\beta$ in Fig. 7 and Fig. 8 graphs (a) were found to be generally smaller, between 0.6 and 0.7. The PNS contraction times exceed more than 1 second, as can be seen from the slowly contracting neutrinospheres in graphs (d). The electron neutrino luminosities in Fig. 7 and Fig. 8 graphs (b) are generally smaller (< 0.5 erg/s) in comparison to the thermal dominated spectra in Fig. 5 and Fig. 6. For low accretion rates, the electron neutrino luminosities are dominated by the accretion luminosity as described by the limiter in Eq. (11).

As an example of an intermediate mass progenitor model with a small mass accretion rate, we apply the same analysis to a core collapse simulation of the 15 M$_\odot$ progenitor from Woosley and Weaver (1995) in Fig. 9 (using *eos2*), which often served as a reference model (e.g. Liebendörfer et al. (2005)). Due to the small mass accretion rate in graph (c), the neutrinospheres in graph (d) contract on timescales of hundreds of milliseconds. In contrast to the massive progenitors with a small mass accretion rate, the neutrino number spectrum at the electron-neutrinosphere differs only slightly from the thermal one and $\beta$ in graph (a) was found to be quite large, between 0.7 increasing up to 0.8 as for the massive progenitors with a large mass accretion rate. The electron-neutrino luminosity in graph (b) agrees initially (until 200 ms post bounce) with the approximation Eq. (11) due to the limiter and is dominated by the diffusion luminosity. Note, although the electron-neutrino Luminosity can be approximated by the accretion luminosity after 200 ms post bounce, the difference from the diffusion luminosity is only $\simeq 10\%$. On a longer timescale, the accretion luminosity becomes too large.

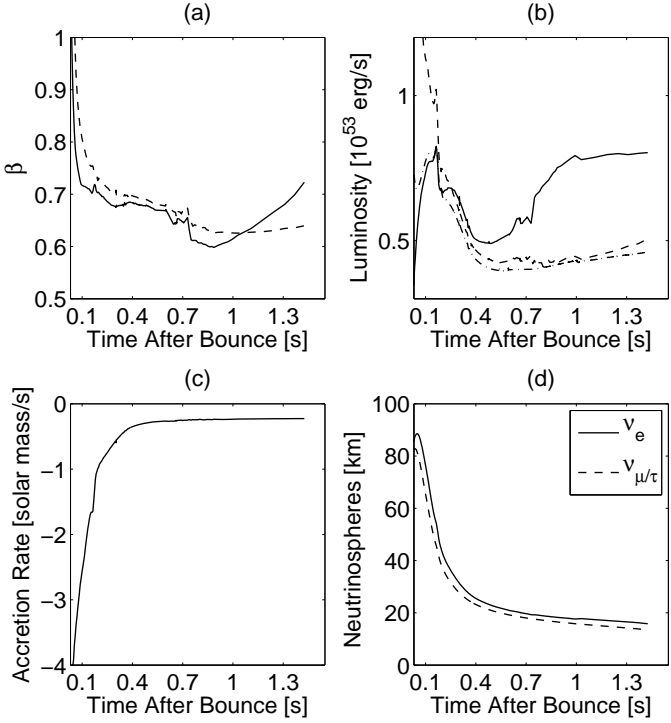
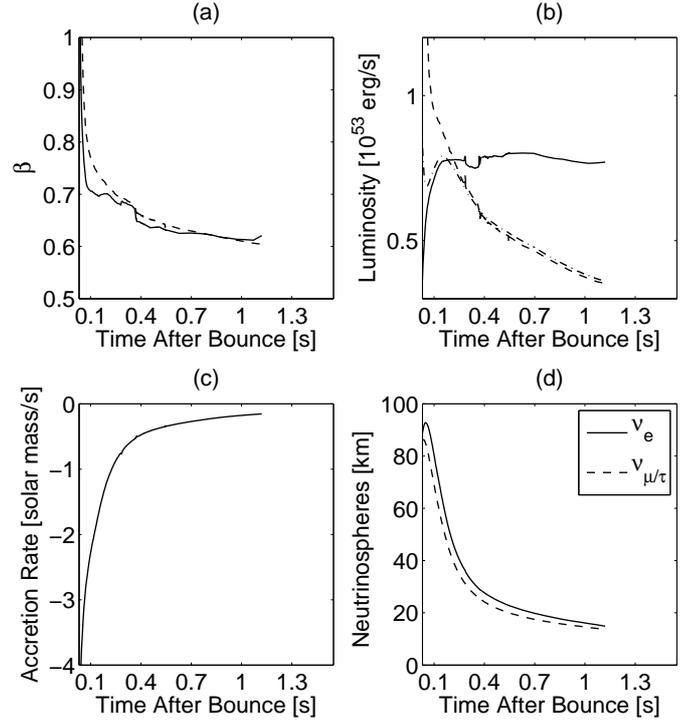

**Figure 7.** The same presentation as Fig. 5 but for a 40 M$_\odot$ progenitor model form Woosley et al. (2002) with a much smaller mass accretion rate. The electron-neutrino luminosity is dominated by the accretion part of Eq. (11).

**Figure 8.** The same presentation as Fig. 7 but for a 50 M$_\odot$ progenitor model from Umeda and Nomoto (2008) with small mass accretion rate. The electron-neutrino luminosity is as well dominated by the accretion part of Eq. (11).

So far, it has been demonstrated that the electron-neutrino luminosity in the post bounce phase of core collapse supernovae depends sensitively on the progenitor model induced neutrino-spherical data and does not generally follow a thermal spectrum.

## 4. Dependency of the emitted neutrino signal on the progenitor model

The shock dynamics during the post bounce evolution of failed core collapse supernova explosions take place in the innermost $\simeq 200$ km in spherical symmetry. Note that present axially symmetric core collapse models are more optimistic. The dynamical evolution outside the Fe-core is determined by the progenitor structure and does not evolve significantly during the simulation time. Material from the gravitationally unstable surrounding regions continues to fall onto the SAS. Nuclei dissociate into free nucleons and light nuclei, which accrete slowly onto the PNS at the center. The PNS contraction is determined by the evolution of the baryon mass

$$\left.\frac{\partial a}{\alpha \partial t}\right|_{R_{\nu_e}} = \frac{4\pi r^2 u \rho}{\Gamma} \qquad (13)$$

at the radius of the electron-neutrinosphere $R_{\nu_e}$. $u = \partial r/\alpha \partial t$ is the radial velocity of the accreting matter, $\rho$ is the matter density, $\Gamma$ is the relativistic factor and $\alpha$ is the lapse function. For the derivation of the expression for the evolution of the gravitational mass, see the appendix of Liebendörfer et al. (2001b).

Above, we have discussed the connection between the emitted neutrino spectra and the matter properties at the neutrinospheres. In the following, we will explore whether there is a correlation between the different neutrino spectra and the structure of the progenitor.

The progenitor models under investigation are the 40 M$_\odot$ from Woosley and Weaver (1995) (40WW95), the 40 M$_\odot$ from Woosley et al. (2002) (40W02), the 40 M$_\odot$ from Umeda and Nomoto (2008) (40U08), the 50 M$_\odot$ from Umeda and Nomoto (2008) (50U08) and the 50 M$_\odot$ from Tominaga et al. (2007); Umeda and Nomoto (2005) (50T07). All are non-rotating and of solar metallicity. These models differ in the size of the iron-cores (see Tab. 2). The masses of the iron-cores are thereby determined intuitively, as Fe-group nuclei ($^{52}$Fe, $^{53}$Fe, $^{56}$Fe and $^{56}$Ni) are more abundant then $^{28}$Si and $^{32}$S. Fig. 10 and Fig. 11 illustrate the post bounce luminosities and mean neutrino energies of the progenitor models listed in Tab. 2. As discussed above, the different mass accretion rates result in different PNS contraction timescales and different electron-neutrino luminosity dependencies.

The models 40WW95, 40U08 and 50T07 with large mass accretion rates at the PNS surface are identified with a short accretion phase after bounce before black hole formation. This corresponds to large luminosities ($L_{\nu_e} > 0.5 - 1$ erg/s, $L_{\mu/\tau} > 0.4$ erg/s) in the graphs (a) and (c). The $(\mu/\tau)$-(anti)neutrino mean energies in the graphs (c) increase rapidly after bounce and reach 34 MeV. On the other hand, the models 40W02 and 50U08 with small mass accretion rates at the neutrinospheres show before black hole formation an extended accretion phase of more than 1 second with smaller luminosities after the neutrino burst has been launched ($L_{\nu_e} < 0.5$ erg/s, $L_{\mu/\tau} < 0.2$ erg/s). The electron-neutrino flavor mean energies in the graphs (b) and (d) follow a similar behavior for all progenitor models while the $(\mu/\tau)$-(anti)neutrino energies increase over a longer timescale during the PNS contraction. For the model 40W02, the $(\mu/\tau)$-(anti)neutrino energies increase only after 700 ms after bounce from 22 MeV up to only 30 MeV. For the model 50U08, the PNS does not reach equivalent compactness during the post bounce accretion phase before becoming gravitationally unstable and collapsing to a black hole. The $(\mu/\tau)$-(anti)neutrino energies increase from 22 MeV to 24 MeV only after 1.1 s after bounce. The progenitor models under investigation are based on different stellar evolution models. They involve different treatments of, for instance, nuclear burning, mixing, neutrino losses and the EoS. Hence, all models show differences at the final stage of stel-lar evolution, as illustrated in Fig. 12 for the 40 $M_\odot$ progenitor models under investigation and Fig. 13 for the 50 $M_\odot$ progenitor models under investigation. To be able to compare the progenitors, we evolve each model until the same central densities are reached. These are $9.12 \times 10^9$ g/cm$^3$ for the 40 $M_\odot$ and $1.58 \times 10^{10}$ g/cm$^3$ for the 50 $M_\odot$ progenitor models.

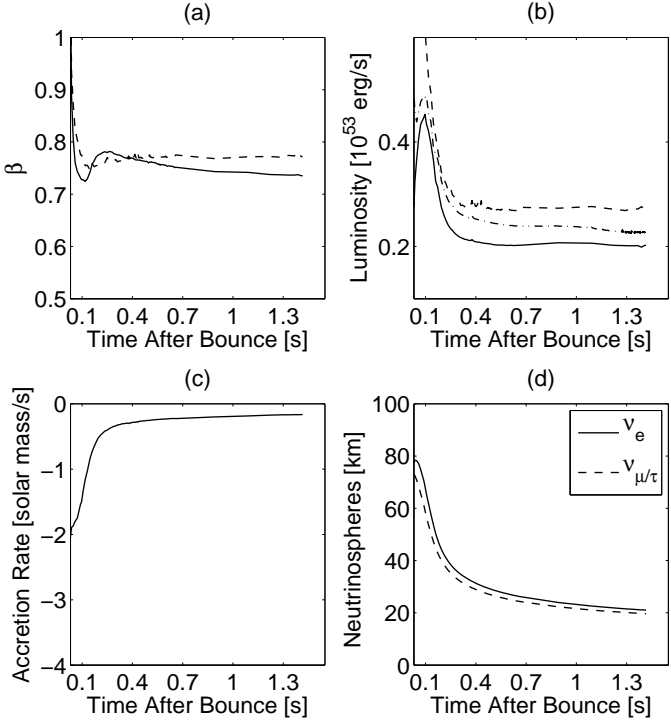

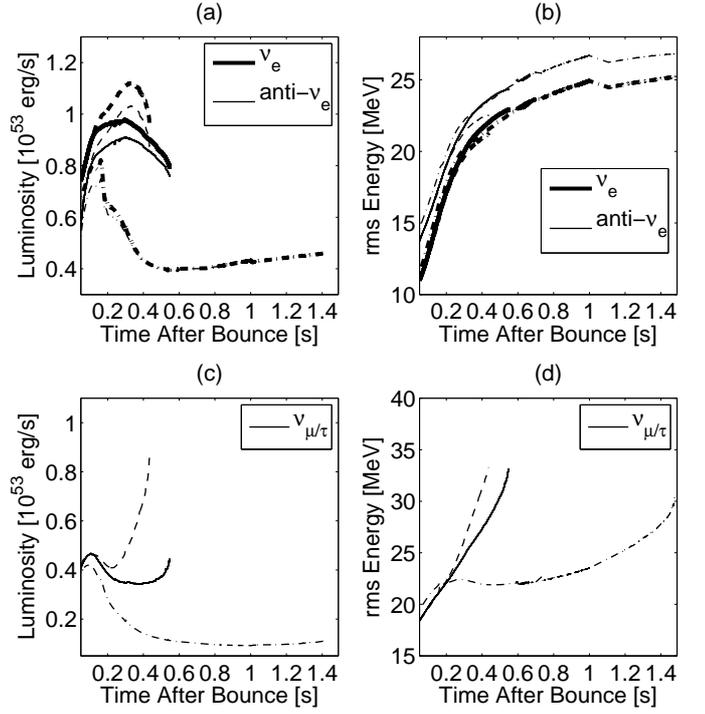

**Figure 9.** The same presentation as Fig. 8 for a 15 $M_\odot$ progenitor model from Woosley and Weaver (1995) with a small mass accretion rate. In contrast to the massive progenitors with a comparable mass accretion, the electron-neutrino luminosity is dominated by the diffusion part of Eq. (11).

**Figure 10.** Neutrino luminosities in the graphs (a) and (c) and mean neutrino energies in the graph (b) and (d) for the different 40 $M_\odot$ progenitor models under investigation (solid lines: 40U08, dashed lines: 40WW95, dash-dotted lines: 40W02) with respect to time after bounce.

**Table 2.** The size of the iron core and time between bounce and black hole formation for the different progenitor models.

| Model | Fe-core [$M_\odot$] | $t_{bh}$ [ms] |
|---|---|---|
| 40WW95 | 1.76 | 435.5 |
| 40W02 | 1.56 | 1476.5 |
| 40U08 | 1.74 | 548.4 |
| 50U08 | 1.89 | 1147.6 |
| 50T07 | Unknown | 487.3 |

Consequently, the central region of all models are very similar. The 40 $M_\odot$ progenitors have electron factions of $Y_e \simeq 0.44$, temperatures of $\simeq 0.85$ MeV and infall velocities of $\sim 1000$ km/s. The more massive 50 $M_\odot$ progenitors have similar infall velocities but higher central temperatures of 0.9 MeV and are slightly more deleptonized with $Y_e \simeq 0.43$. Note that the central hydrodynamical variables are rather similar compared to the properties outside the Fe-cores (see Fig. 12 and Fig. 13). There, the differences of the baryon density can be more than one order of magnitude for the same progenitor mass while temperatures and infall velocities can differ by a factor of 2. These differences are responsible for the different dynamical evolution in the post bounce phase and will be discussed in the following paragraph.

The central regions evolve in a similar manner and are synchronized at core bounce for all progenitor models, as illustrated in Fig. 14 and Fig. 15. As demonstrated above, the PNS contraction behavior and the subsequent electron-neutrino luminosity are determined by the mass accretion rate at the neutrinosphere, which depends on the amount of mass that falls through the SAS from regions outside the Fe-core. Hence, the mass accretion rate at the neutrinosphere is given by the progenitor structure at bounce as illustrated in Fig. 14 and Fig. 15.

The infall velocities in Fig. 14 and Fig. 15 (a) are similar for all progenitor models. On the other hand, for the same progenitor mass differ the matter densities in the graphs (b) outside the Fe-cores substantially, comparing the models from the different stellar evolution groups.

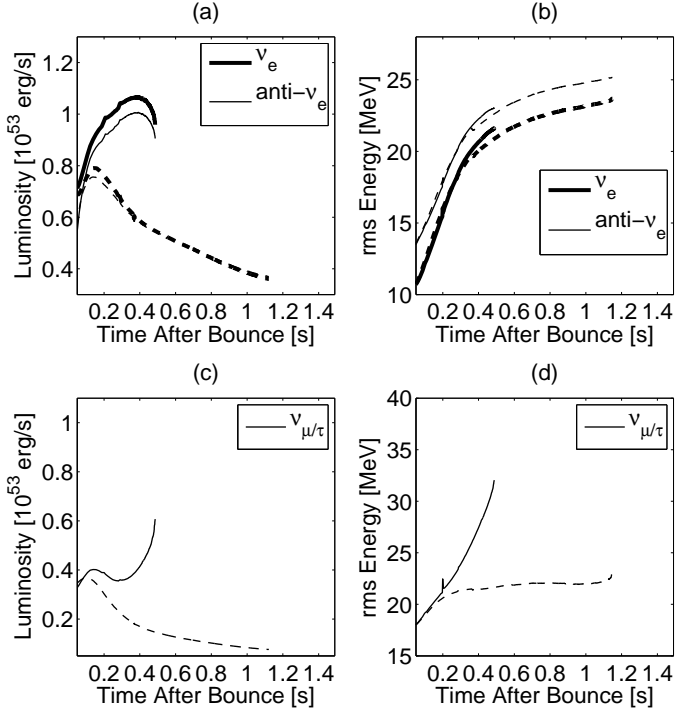

**Figure 11.** The same presentation as Fig. 10 for the different 50 M$_\odot$ progenitor models under investigation (solid lines: 50T07, dashed lines: 50U08).

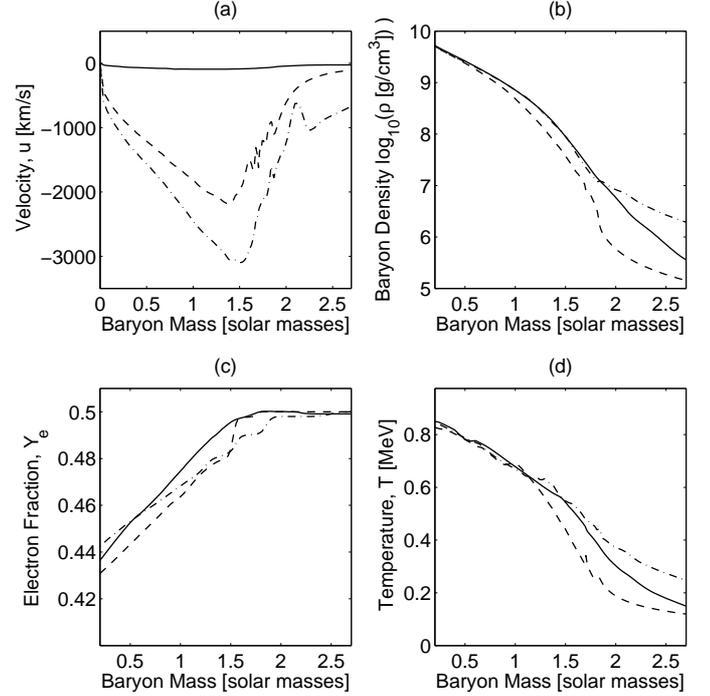

**Figure 12.** Selected hydrodynamic variables for the different 40 M$_\odot$ progenitor models under investigation (solid lines: 40U08, dashed lines: 40WW95 and dash-dotted lines: 40W02) at the final stage of stellar evolution.

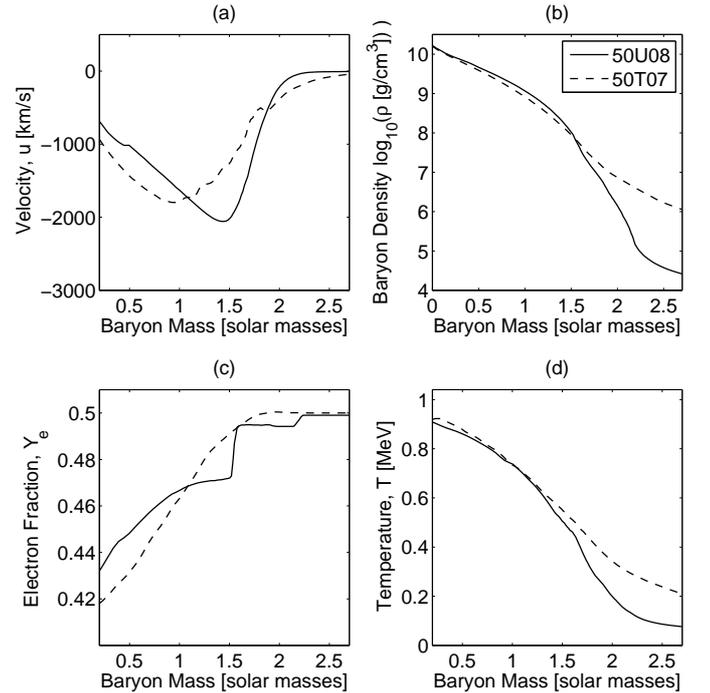

**Figure 13.** The same configuration as Fig. 12 for the different 50 M$_\odot$ progenitor models under investigation (solid lines: 50U08, dashed lines: 50T07).

The models 40WW95, 40U08 and 50T07 have high matter densities outside the Fe-cores. This corresponds to large mass accretion rates at the neutrinospheres and short post bounce accretion times before the PNSs reach the critical masses and collapse to black holes (see Fig.10 and Fig.11). The electron neutrino luminosities correspond to thermal spectra.

The opposite holds for the models 40W02 and 50U08, with small matter densities outside the Fe-cores. This leads to small mass accretion rates at the neutrinospheres and consequently extended post bounce accretion phases (see Fig.10 and Fig.11). The electron neutrino luminosities correspond to accretion spectra.

We have found a correlation between the electron flavor neutrino luminosities and the progenitor structure. The latter has a direct impact on the mass accretion rate at the neutrinosphere and hence on the electron neutrino spectra. We find that the structure of the progenitor has a non-negligible influence on the emitted neutrino spectra. This is in contradiction to Sumiyoshi et al. (2008), who attribute differences in the emitted neutrino emission mainly to the properties of the EoS. This is due to their selective choice of progenitor models, which all have large accretion rates producing thermal electron-neutrino spectra. However, for models that have been used in both studies, e.g. 40WW95 and 50T07, the results are qualitatively similar.

## 5. Neutral current reactions in core collapse supernovae

The standard neutrino emissivities and opacities as well as the scattering kernels are given in Bruenn (1985). The dominant reactions for the electron-flavor neutrinos are the charged current reactions. The source for $(\mu/\tau)$-neutrinos are pair-processes, which are assumed to interact only via neutral-current reactions. Focusing on the pair-processes here, attention is devoted to the $(\mu/\tau)$-neutrinos.

The classical pair-production process is

$$e^- + e^+ \rightleftarrows \nu_i + \bar{\nu}_i. \tag{14}$$

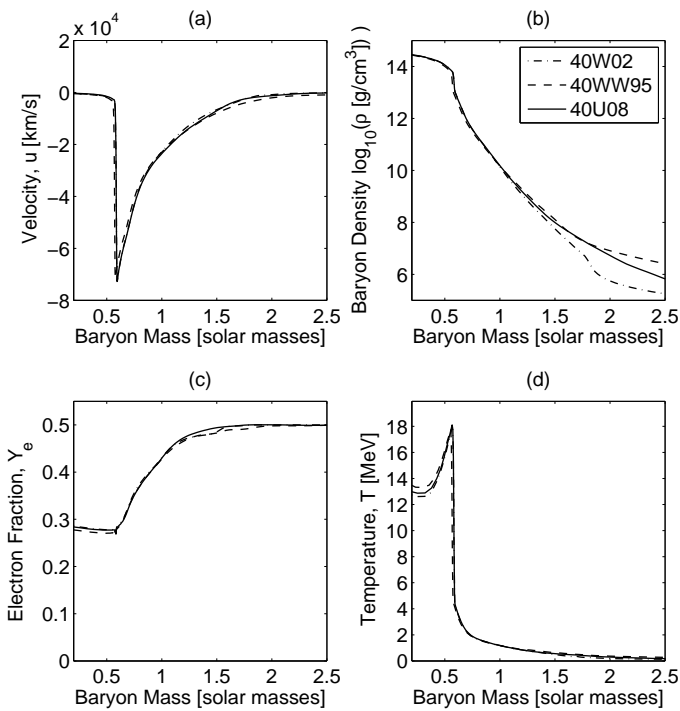

**Figure 14.** Bounce conditions for the 40 $M_\odot$ progenitor models under investigation

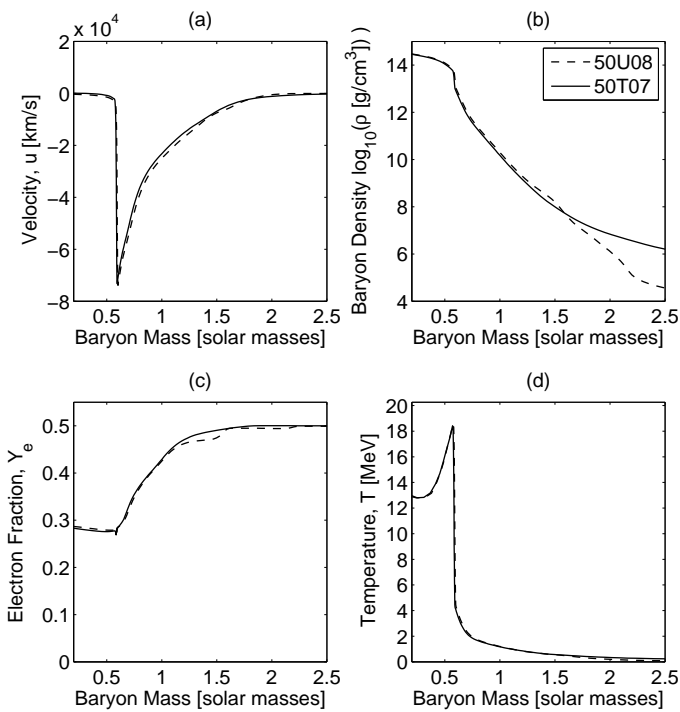

**Figure 15.** Bounce conditions for the 50 $M_\odot$ progenitor models under investigation

Additional pair emission and thermalisation processes have been investigated by Thompson and Burrows (2001), such as Nucleon-Nucleon-Bremsstrahlung

$$N + N \rightleftarrows N' + N' + \nu_i + \bar{\nu}_i, \qquad (15)$$

(with $i \in (e, \mu, \tau)$), which has been included into our model by Messer and Bruenn (2003) based on Hannestad and Raffelt (1998). Improvements of these rates have recently been published in Bacca et al.. Buras et al. (2003) investigated further $(\mu/\tau)$-(anti)neutrino pair emission via the annihilation of trapped electron neutrino and anti-neutrino pairs

$$\nu_e + \bar{\nu}_e \rightleftarrows \nu_{\mu/\tau} + \bar{\nu}_{\mu/\tau}. \qquad (16)$$

We show in the Appendix that the reaction rates for (14) and (16) can be calculated among similar lines, via phase space integrations over the distribution functions for incoming and outgoing particles as well as the squared and spin averaged matrix element $\langle M \rangle^2$. Note that the favoured region for $(\mu/\tau)$-neutrino pair emission and absorption via reaction (16) lies inside the electron-neutrinosphere, where electron-(anti)neutrinos are in local thermodynamic equilibrium (LTE) with matter via the electronic charged current reactions

$$e^- + p \leftrightarrows n + \nu_e, \qquad (17)$$

$$e^+ + n \leftrightarrows p + \bar{\nu}_e. \qquad (18)$$

Here, we compare the different neutral current reactions (14)-(16).

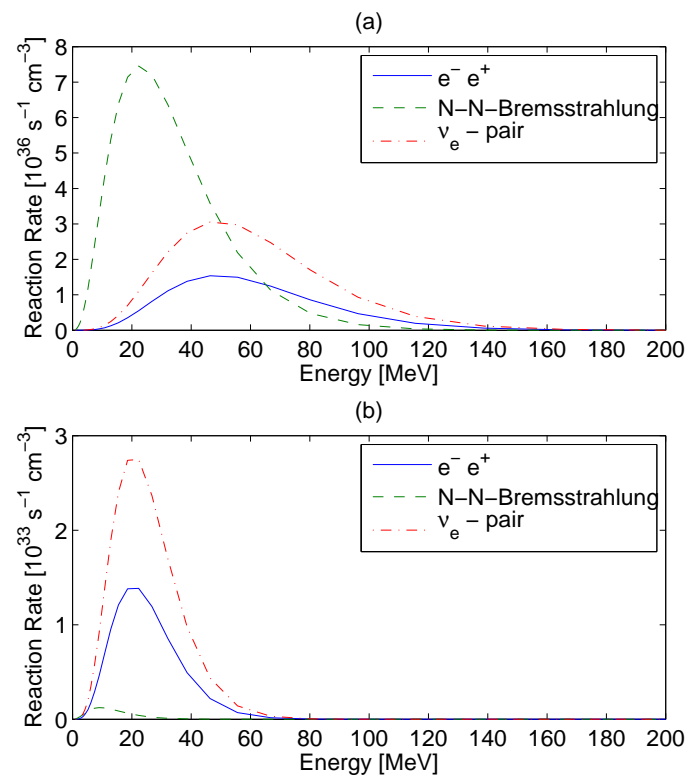

**Figure 16.** The different $(\mu/\tau)$-neutrino-pair reaction rates (14), (15) and (16) as a function of the neutrino energy for the two different hydrodynamical conditions (a: $T = 12$ MeV, $\rho = 5 \times 10^{13}$ g/cm$^3$, $Y_e = 0.3$) and (b: $T = 5$ MeV, $\rho = 1 \times 10^{12}$ g/cm$^3$, $Y_e = 0.3$), at zero chemical potentials ($\mu_e = 0$ and $\mu_{\nu_e} = 0$).

Fig. 16 illustrates the three phase-space-integrated reaction rates separately at two different hydrodynamical states, as a function of the neutrino energy. At high matter density and temperature in graph (a), N-N-Bremsstrahlung is found to be the dominant reaction for the emission of $(\mu/\tau)$-(anti)neutrinos. This is in agreement with an earlier study by Messer and Bruenn (2003). Following the path to lower densities and temperatures in graph (b) but still inside the $\nu_e$-sphere, in agreement with Buras et al. (2003) we find reaction (16) dominates the other pair-production rates (14) and (15). However, outside the $\nu_e$-sphere, reaction (16) no longer contributes to the emission of $(\mu/\tau)$-(anti)neutrino pairs since matter becomes more and more transparent to neutrinos. The dominant source for pair-reactions are the electron-positron pairs.

After the separate investigation of the pair reactions, we will now explore the effects observed in core collapse simulations comparing the different sets of pair-reactions (14) and (15) with the full set of reactions (14)-(16).

The additional source of $(\mu/\tau)$-(anti)neutrinos via reaction (16) increases the total $(\mu/\tau)$-neutrino pair-reaction rate, which increases the $(\mu/\tau)$-neutrino luminosity in Fig. 17 (a) and the mean neutrino energies in Fig. 17 (b) from regions inside the $\nu_e$-sphere. Note, although N-N-Bremsstrahlung dominates the low neutrino energy regime, its contribution to change the $(\mu/\tau)$-neutrino luminosity is relatively minor.

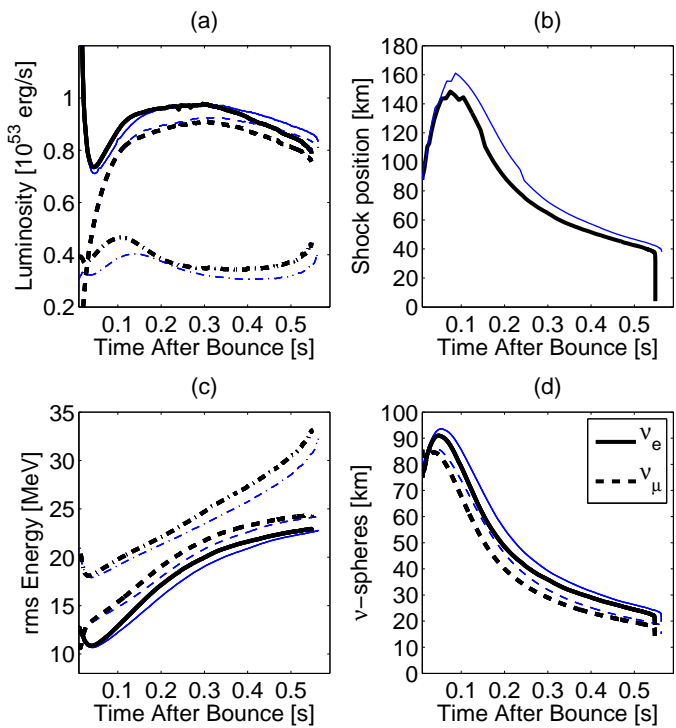

**Figure 17.** The neutrino luminosities in graph (a) and mean neutrino energies in graph (c) of all three neutrino flavors (solid: $\nu_e$, dashed: $\bar{\nu}_e$, dash-dotted: $(\mu/\tau)$-(anti)neutrinos) as a function of time after bounce, comparing the differences of the pair reactions (14) and (15) (thin blue lines) and (14)-(16) (thick black lines) during the post bounce evolution of a 40 $M_\odot$ progenitor from Umeda and Nomoto (2008). In addition, we compare the position of the shocks in graph (b) and the radii of the neutrinospheres in graph (d) as a function of time after bounce.

Since $(\mu/\tau)$-neutrinos do not interact via the charged current reactions, the generally larger $(\mu/\tau)$-neutrino luminosity and mean energies results in larger cooling rates at the $(\mu/\tau)$-neutrinospheres. This leads to a more compact central object as indicated by the shock and neutrinosphere position in Fig. 17 graphs (b) and (d) respectively. The gain region behind the shock does not expand as much and the infall velocities ahead of the shock are larger. Losing pressure support from below, the shock starts to propagate inward earlier. Due to the higher temperatures at the PNS surface, the electron-flavor neutrino mean energies are larger (see Fig. 17 graph (c)). However, the faster contracting neutrinospheres in graph (d) lead to smaller electron-flavor neutrino luminosities after 400 ms after bounce in graph (a). An additional effect observed via the comparison of fast and slow contracting PNSs (achieved via different cooling at the PNS surface) is the shorter accretion time before becoming gravitationally unstable. Here, we find the accretion time to be shorter by $\simeq 10$ ms. Hence, the $(\mu/\tau)$-neutrino reaction and cooling rates, reflect sensitively the matter conditions at the neutrinospheres, the PNS contraction behaviour and the emitted neutrino spectra.

Buras et al. (2003) have seen similar phenomena with respect to the neutrino luminosities and mean energies. They also see the more compact PNS and the less extended gain region behind the shock. However, since their simulations only lasted for $\simeq 350$ ms post bounce and they were investigating a 15 $M_\odot$ progenitor model with a lower accretion rate (in comparison to our 40 $M_\odot$ progenitor), the effects observed were less intense and did not include the formation of a black hole.

### 5.1. Evolution of the $\mu/\tau$-neutrino luminosity

In the following, we extend the analysis of Liebendörfer et al. (2004) and Fischer et al. (2007), who investigated the drastic $(\mu, \tau)$-(anti)neutrino luminosity increase during the late PNS accretion phase of failed core collapse supernova explosions of massive progenitors.

The evolution of the neutrino luminosities depends on the production rates and the diffusion timescale, which in turn depend on the assumed matter conditions. These conditions and the production rates for all neutrino flavors are plotted in Fig. 18 for a core collapse simulation of a 40 $M_\odot$ progenitor model, applying the full set of pair reactions (14)-(16). In order to separate the different regimes, we plot all quantities with respect to the baryon density. Note, that the electron-(anti)neutrinospheres are at lower densities than the $(\mu/\tau)$-neutrinospheres, since the latter do not interact via charged current reactions.

Most $(\mu, \tau)$-(anti)neutrino pairs are produced at $\rho \simeq 10^{13}$ g/cm$^3$. This finding remains rather constant during the late accretion phase, because the matter temperature $T$ and the electron fraction $Y_e$ do not change at that density, as can be seen in Fig. 18 (e) and (f). In contrast, at the $(\mu/\tau)$-neutrinosphere ($\rho \simeq 10^{11}$ g/cm$^3$) we find a drastic increase in temperature and $Y_e$. Due to the continuous contraction of the PNS, the electron-degeneracy reduces which favors more electron-positron-pairs. These thermalized electron-positron-pairs increase the $(\mu/\tau)$-neutrino pair reaction rates in graph (d) via reaction (14), which increases the $(\mu/\tau)$-(anti)neutrino luminosity contribution from lower densities. In addition, the diffusion time scale of the $(\mu/\tau)$-neutrinos is reduced during the PNS contraction. The corresponding optical depths (at 300 km distance) are shown in graph (c).

### 5.2. Improvements of the neutrino opacities

Finally, we will investigate corrections of the standard neutrino opacities (Bruenn (1985)), following the suggestions by Horowitz (2002) regarding the effect of weak magnetism, nucleon recoil and corrections for the strangeness of nucleons, as already briefly explored in Liebendörfer et al. (2003) using a 15 $M_\odot$ progenitor model. We will illustrate the effects using the example of a failed supernova explosion of a 40 $M_\odot$ progenitor model from Umeda and Nomoto (2008) during the post bounce evolution.

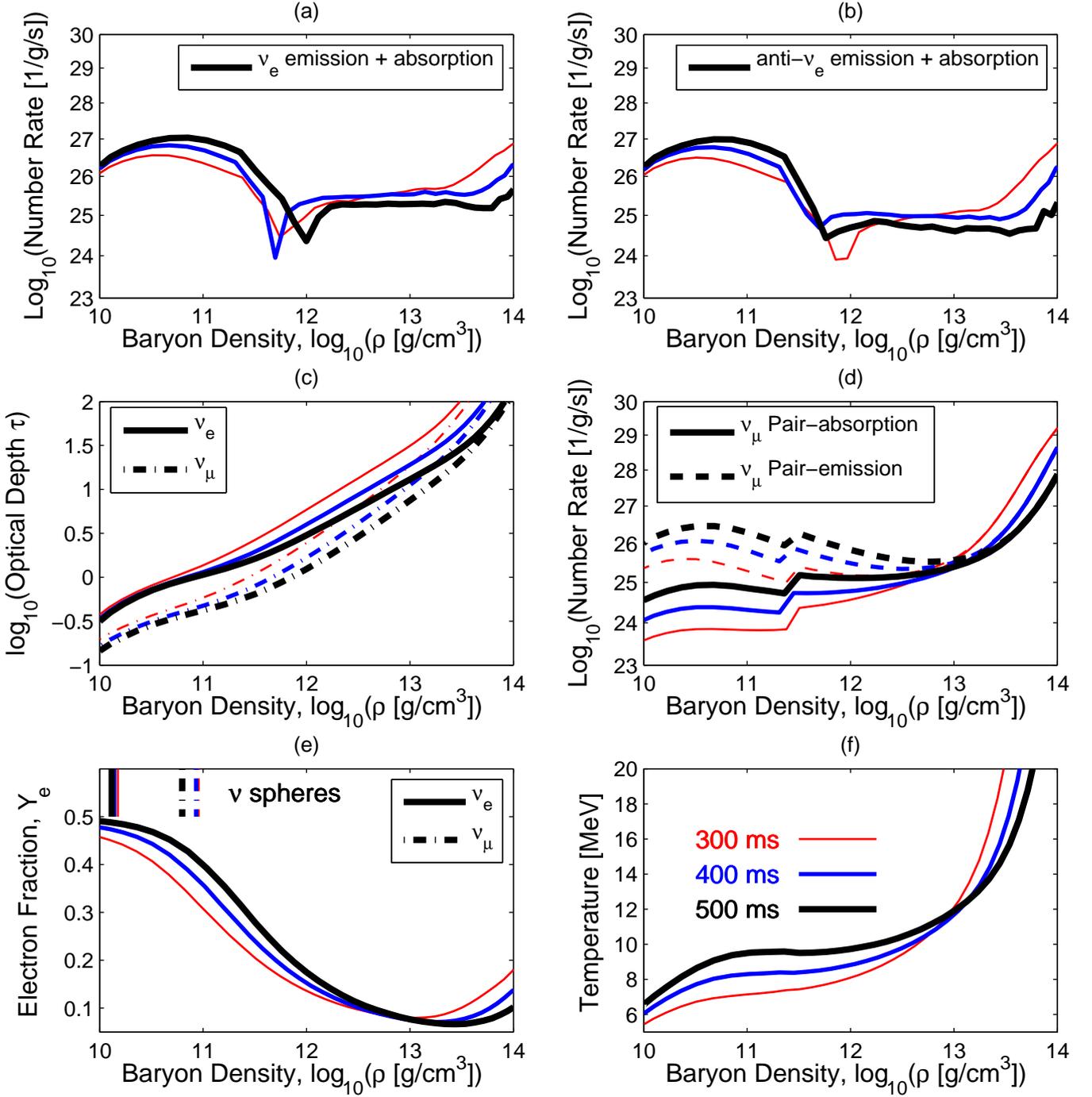

**Figure 18.** Different post bounce snap shots (thin red lines: 300 ms intermediate blue lines: 400 ms thick black lines: 500 ms), illustrating the effects of the PNS contraction to the number fluxes, the optical depths $\tau$, the matter temperature and the electron fraction $Y_e$.

The separate consideration of weak magnetism is a higher order extension of the zeroth order scattering cross section, which reduces the antineutrino and increases the neutrino cross sections. On the other hand, recoil reduces both neutrino- and antineutrino cross sections, as discussed in Horowitz (2002). The total modified cross section for the electronic charged current reactions

$$\nu_e + n \rightarrow p + e^-,$$
$$\bar{\nu}_e + p \rightarrow n + e^+,$$

can be written as

$$\sigma = \sigma_0 R(E),$$

with zeroth order cross section $\sigma_0$. $R$ depends on the neutrino energy $E$. For average neutrino energies, $R \simeq 1$ for the electron neutrinos, and $R < 1$ for the electron-antineutrinos. As discussed in Horowitz (2002), this effect reduces the electron-antineutrino opacity from regions inside the neutrinosphere, as illustrated in Fig. 19 resulting in a larger electron-antineutrino luminosity in graph (a) and larger mean neutrino energies in graph (b). The electron neutrino luminosity and mean neutrino energies remain almost unaffected. This phenomenon becomes important during the PNS accretion phase, as the matter density rises and more neutrinos are found at higher energies, where the higher order corrections become significant. The electron-antineutrino luminosity rises above the electron neutrino luminosity, here after 400 ms post bounce, as the PNS contraction reaches a certain level of compactness.

The correction from the strange quark contributions are taken into account by a modified axial-vector coupling constant, for (anti)neutrino-nucleon scattering

$$\nu + N \to \nu + N.$$

The larger electron-antineutrino cooling rates inside the neutrinosphere result in a more compact PNS supporting higher matter temperatures, compared to core collapse simulations of the same progenitor model with otherwise identical input physics. The largest differences are found at the ($\mu/\tau$)-neutrinosphere, while the contribution from lower matter densities remains unaffected. The higher matter temperatures are directly reflected in larger neutrino-pair reaction rates at regions near the ($\mu/\tau$)-neutrinosphere, while the contribution from high matter density at $\rho \sim 10^{13}$ g/cm$^3$ remains constant. The larger ($\mu/\tau$)-neutrino pair reaction rate increases the ($\mu/\tau$)-neutrino luminosity from regions with intermediate matter density $\rho \sim 10^{11} - 10^{12}$ g/cm$^3$. This increases the effect of cooling, which supports a more compact PNS. The more rapidly contracting PNS at the center consequently leads to a shorter PNS accretion time, before becoming gravitationally unstable and collapsing to a black hole. For the model considered here, the difference is $\simeq 20$ ms as can be seen in Fig. 19.

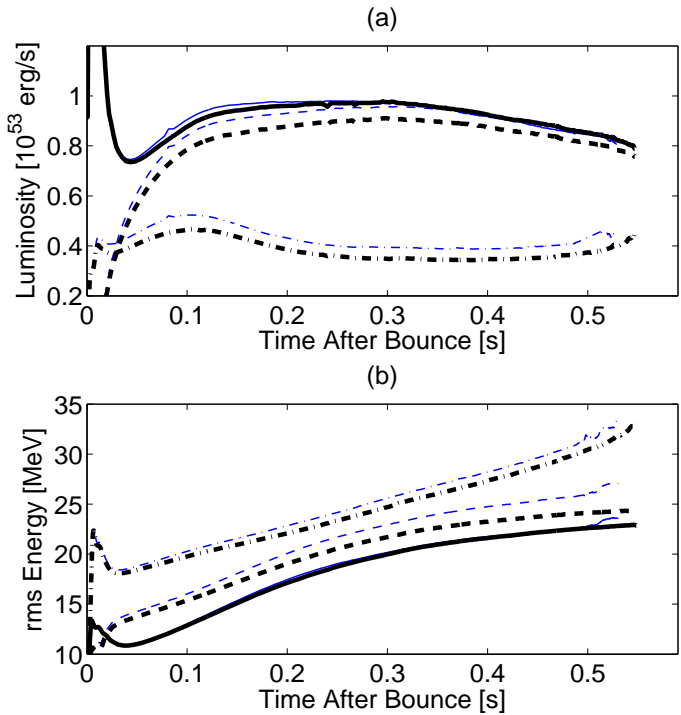

**Figure 19.** Comparing the standard neutrino opacities (thick black lines) (see Bruenn (1985)) with the corrections (thin blue lines) given in Horowitz (2002), plotting the neutrino luminosities and the mean neutrino energies as a function of time after bounce, for all three neutrino flavors (solid line: $\nu_e$, dashed: $\bar{\nu}_e$, dash-dotted: $\mu/\tau$-neutrinos) at 500 km distance.

## 6. Summary

We performed spherically symmetric core collapse simulations of massive progenitors using general relativistic radiation-hydrodynamics with three-flavor Boltzmann neutrino transport. In the absence of an earlier explosion, the continuous accretion of material onto the central PNS will eventually lead to the formation of a solar mass black hole, on timescales up to seconds for progenitors in the mass range of 40 to 50 M$_\odot$.

The neutrinos emitted during core collapse supernovae are, besides gravitational waves and nucleosynthesis yields, the only source of information leaving the stellar core. In addition, the available NS properties from observations provide information about the remnants of core collapse supernova explosions. However, gravitational waves are difficult to detect, nucleosynthesis calculations are model dependent and NS mass measurements provide information about the EoS of hot and dense nuclear matter. Hence, we believe that up to now neutrinos are the most promising source of information that gives a direct insight into the happenings inside the stellar core. The understanding of the emission, absorption and transport of neutrinos is essential for the accurate modeling of core collapse supernovae. Special focus is devoted to the cooling at the neutrinospheres and heating between the neutrinospheres and the expanding shock during the post bounce evolution.

We confirm the results from Sumiyoshi et al. (2007), that a stiff EoS for hot and dense nuclear matter leads to an extended accretion phase of several seconds (as has been explored here at the example of a 40 M$_\odot$ progenitor model).

However, comparing progenitors of the same mass from different stellar evolution groups shows that a small mass accretion rate at the electron-neutrinosphere also leads to an extend accretion phase of several seconds. For the same progenitor mass but different mass accretion rates at the electron-neutrinosphere, we even find a different electron-neutrino luminosity dependency. Models with large mass accretion are determined by a diffusion dominated electron-neutrino spectrum, while small mass accretion rates lead to accretion dominated spectra. Different EoS of hot and dense nuclear matter are unable to change a diffusion dominated electron-neutrino spectrum into an accretion dominated one. Different EoSs might extend or delay the accretion phase due to a different compressibility and asymmetry energy or may provide a different composition. However, the electron-neutrino spectra will always stay either diffusion or accretion dominated, determined by the progenitor model only. In that sense, the progenitor model has a non-negligible influence on the emitted neutrino spectra. This is in contradiction to the recently published work by Sumiyoshi et al. (2008). They investigated different progenitor models with similar mass accretion rates and thus find the progenitor dependency less relevant for the emitted neutrino signal. We would like to point out that the emitted neutrino signal contains correlated information about the EoS, the progenitor star and the neutrino physics. If analyzing the neutrino luminosities, one has to take all these dependencies into account.

Finally, three-dimensional core collapse models have to make use of some from of neutrino approximation scheme due to present computational limitations. For that reason, we introduced an electron neutrino luminosity approximation which can be applied to any progenitor model and for large distances, typically from a few 100 km to the remaining physical domain of the progenitor. This approximation depends only on the mass accretion rate, given by the progenitor model, and the temperature at the electron-neutrinosphere. We compared this approximation with accurate three-flavor Boltzmann neutrino transport calculations for several different massive progenitor models and find qualitatively good agreement.

**Table 3.** Weak interaction coefficients in the Weinberg-Salam-Glashow theory (in first order), where $\langle M \rangle^2$ is the squared and spin-averaged matrix element and $\theta_W$ is the Weinberg angle. $\mathbf{p}_i$ are the 4-momenta of the interacting particles. For the calculation of the proper phase space integration with the assumption of a homogeneous distribution as well as various additional neutrino interaction processes, see for example Hannestad and Madsen (1995).

| Reaction | $\langle M \rangle^2$ | $c_V$ | $c_A$ |
|---|---|---|---|
| $\nu_i + e^- \leftrightarrows \nu'_i + e'^-$ | $32G_F^2((c_A+c_V)^2(\mathbf{p}_\nu \cdot \mathbf{p}_e)(\mathbf{p}_{\nu'} \cdot \mathbf{p}_{e'})$ $+(c_A-c_V)^2(\mathbf{p}_\nu \cdot \mathbf{p}_{e'})(\mathbf{p}_e \cdot \mathbf{p}_{\nu'}))$ $-(c_V^2-c_A^2)^2 m_e^2 (\mathbf{p}_\nu \cdot \mathbf{p}_{\nu'}))$ | $\frac{1}{2} + 2\sin\theta_W$ | $\frac{1}{2}$ |
| $e^- + e^+ \leftrightarrows \nu_i + \bar{\nu}_i$ | $32G_F^2((c_A+c_V)^2(\mathbf{p}_{\nu_i} \cdot \mathbf{p}_{e^+})(\mathbf{p}_{\bar{\nu}_i} \cdot \mathbf{p}_{e^-})$ $+(c_A-c_V)^2(\mathbf{p}_{\nu_i} \cdot \mathbf{p}_{e^-})(\mathbf{p}_{\bar{\nu}_i} \cdot \mathbf{p}_{e^+}))$ $-(c_V^2-c_A^2)^2 m_e^2 (\mathbf{p}_{\nu_i} \cdot \mathbf{p}_{\bar{\nu}_i}))$ | $\frac{1}{2} + 2\sin\theta_W$ | $\frac{1}{2}$ |
| $\nu_e + \bar{\nu}_e \leftrightarrows \nu_{\mu,\tau} + \bar{\nu}_{\nu,\tau}$ | $32G_F^2(\mathbf{p}_{\nu_{\mu,\tau}} \cdot \mathbf{p}_{\bar{\nu}_e})(\mathbf{p}_{\bar{\nu}_{\mu,\tau}} \cdot \mathbf{p}_{\nu_e})$ | $\frac{1}{2}$ | $\frac{1}{2}$ |

In addition to the electron-flavor neutrino spectra, the $(\mu/\tau)$-neutrinos are analyzed in the full Boltzmann model and their importance with respect to cooling at the $(\mu/\tau)$-neutrinosphere is explored. We compared different $(\mu/\tau)$-neutrino pair reactions separately and during the accretion phase of failed core collapse supernova explosions of massive progenitors. A large $(\mu/\tau)$-neutrino luminosity corresponds to a large cooling rate and consequently supports a more compact PNS as well as a shorter PNS accretion phase of the order of a few milliseconds. In addition, the connection between the drastic $(\mu/\tau)$-neutrino luminosity increase during the accretion phase and the PNS contraction has been investigated. We find that the changing thermodynamic conditions (especially the increasing temperature) at the $(\mu/\tau)$-neutrinosphere establish $\beta$-equilibrium at a larger value of the electron fraction, which leads to an increase of the $(\mu/\tau)$-neutrino pair reaction rate. This increases the $(\mu/\tau)$-neutrino luminosity from regions with low and intermediate density. Additionally, following Horowitz (2002), an update of the standard neutrino emissivities and opacities has been investigated, using the example of a failed core collapse supernova of a massive progenitor.

The present analysis addresses a deeper understanding of the origin and the dependencies of the emitted neutrino signal of all three flavors, examined using failed core collapse supernova explosions of massive stars and the formation of solar mass black holes.

## Acknowledgment

We would like to thank Simon Scheidegger, Roger Kaeppeli and Urs Frischknecht for many helpful discussions and guiding hints. The project was funded by the Swiss National Science Foundation grant. no. PP002-106627/1 and 200020-122287. A.Mezzacappa is supported at the Oak Ridge National Laboratory, which is managed by UT-Battelle, LLC for the U.S. Department of Energy under contract DE-AC05-00OR22725.

## Appendix

The neutrino pair-emissivity from electron-positron annihilation was calculated by Yueh and Buchler (1976a) and Yueh and Buchler (1976b). They explored the possibility of calculating the pair-reaction rates in a similar way as neutrino-electron scattering (NES). The rate for NES is given by the following integral expression over the phase space distribution functions for incoming $f(\varepsilon)$ and outgoing $f(\varepsilon')$ electrons (including blocking factors) and the squared and spin-averaged matrix element $\langle M \rangle^2$ (listed in Tab. 3)

$$R_{\nu e} = \frac{2}{(2E)} \int \frac{d^3 p}{(2\pi)^3 (2\varepsilon)} f(\varepsilon)(1 - f(\varepsilon')) \int \frac{d^3 q'}{2E'}$$
$$\times \int \frac{d^3 p'}{2\varepsilon'} \left( \frac{\langle M \rangle^2}{(2\pi)^2} \right) \delta^4 (\mathbf{p}_e + \mathbf{p}_\nu - \mathbf{p}'_e - \mathbf{p}'_\nu). \quad (19)$$

The 4-momenta of the in- and outgoing electrons are denoted by $\mathbf{p_e} = (p, \varepsilon)$, $\mathbf{p'_e} = (p', \varepsilon')$ and the scattering neutrinos are denoted by $\mathbf{p_\nu} = (q, E)$, $\mathbf{p'_\nu} = (q', E')$ respectively. Following the argumentation and derivation by Yueh and Buchler (1976b), the emissivity for the traditional neutrino pair reaction (14) is given by the following expression

$$j_\nu = 2\pi E^2 \int \frac{2d^3 p}{(2\pi)^3 (2\varepsilon)} \int \frac{2d^3 p'}{2'} f_{e^+}(\varepsilon) f_{e^-}(\varepsilon')$$
$$\times \left( \frac{\langle M \rangle^2}{(2\pi)^2} \right) \delta^4 (\mathbf{p}_\nu + \mathbf{p}_{\bar{\nu}} - \mathbf{p}_{e^-} - \mathbf{p}_{e^+}) \int \frac{d^3 q'}{2E'}, \quad (20)$$

where $\mathbf{p}_{e^-} = (p', \varepsilon')$, $\mathbf{p}_{e^+} = (p, \varepsilon)$ are the incoming electron/positron 4-momenta and $\mathbf{p}_\nu = (q', E')$, $\mathbf{p}_{\bar{\nu}} = (q, E)$ are the outgoing neutrino 4-momenta. $j_\nu$ differs from $R_{\nu e}$ by the additional multiplicative term $E^3/(2\pi)^3$ and the replacement of the blocking factor $1 - f(\varepsilon')$ by $f_{e^-}(\varepsilon')$. Note that by calculating the reverse reaction rate (the amount of neutrinos absorbed by reaction (14)), the phase-space distribution functions $f_{e^-}(\varepsilon')$ and $f_{e^+}(\varepsilon)$ have to be replaced by the blocking factors $1 - f_{e^-}(\varepsilon')$ and $1 - f_{e^+}(\varepsilon)$ respectively.

Evaluating the squared and spin-averaged matrix element given in Tab. 3, the expression (20) can be rewritten as [see Schinder and Shapiro (1982)]

$$j_\nu = \frac{\sigma}{2} \frac{1}{EE'} \left( (c_A + c_V)^2 I_1 + (c_A - c_V)^2 I_2 + (c_V^2 - c_A^2)^2 I_3 \right), \quad (21)$$

with $\sigma = \frac{1}{(2\pi)^5} \frac{(2G\hbar)^2}{c^5}$. The terms $I_i(E, E', \omega)$ are the phase-space integrals over the distribution functions as well as the separate 4-momenta terms of the squared and spin averaged matrix elements (expressed in terms of Fermi-Dirac integrals depending on the incoming $E$ and the outgoing $E'$ neutrino energies, the electron mass $m_e$, the temperature and the electron chemical potential $\mu_e$), which are given in for example Yueh and Buchler (1976b), Schinder and Shapiro (1982) and Bruenn (1985). The weak interaction coefficients $c_A$ and $c_V$ are listed in Tab. 3 for the different reactions. $\omega$ denotes the relative angle between the incoming $\mu$ and outgoing $\mu'$ neutrino propagation direction and $\phi$ is the azimuthal angle between the two directions, given by the following relation

$$\omega = \mu\mu' + \sqrt{(1-\mu^2)(1-\mu'^2)}\cos\phi. \quad (22)$$

Now, replacing the incoming electron and positron distributions in expression (20) by electron-(anti)neutrino distributions and comparing the weak interaction coefficients of reaction (14) with reaction (16) as well as the corresponding matrix elements in Tab. 3, the pair-reaction rate (21) reduces to

$$j_{\nu_\mu}(E,E',\omega) = \frac{\sigma}{2}\frac{1}{EE'}\widetilde{I}_1(E,E',\omega), \quad (23)$$

where the emitted particles are now $(\mu,\tau)$-neutrino-antineutrino pairs. The term $\widetilde{I}_1$ differs from $I_1$ by the fact that all rest mass dependency has been canceled, since the neutrino masses are assumed to be negligible. The corresponding terms for $I_2$ and $I_3$ do not appear any longer. The electron and positron chemical potentials are replaced by the electron-(anti)neutrino chemical potentials $\mu_{\nu_e} = \mu_e - (\mu_n - \mu_p)$ and $\mu_{\bar{\nu}_e} = \mu_{\nu_e}$ respectively, with proton $\mu_p$ and neutron chemical potential $\mu_n$, since the electron-(anti)neutrinos are in LTE via reactions (17) and (18).